\documentclass[nofootinbib,showpacs,showkeys,prd]{revtex4}
\usepackage{amsmath}
\usepackage{amsfonts}
\usepackage{amssymb}
\usepackage{graphicx}
\usepackage{color}

\begin{document}

\title{Decoherence, Entanglement and Cosmic Evolution}

\author{Salvatore Capozziello}
\email{capozzie@na.infn.it}
\affiliation{Dipartimento di Fisica, Universit\`a di Napoli "Federico II",
             Via Cinthia, I-80126, Napoli, Italy}
\affiliation{INFN Sez. di Napoli, Compl. Univ. Monte S. Angelo Ed. N
Via Cinthia, I- 80126 Napoli, Italy.}

\author{Orlando Luongo}
\email{luongo@na.infn.it}
\affiliation{Dipartimento di Fisica, Universit\`a di Napoli "Federico II",
             Via Cinthia, I-80126, Napoli, Italy}
\affiliation{INFN Sez. di Napoli, Compl. Univ. Monte S. Angelo Ed. N
Via Cinthia, I- 80126, Napoli, Italy.}
\affiliation{Instituto de Ciencias Nucleares, Universidad Nacional Aut\'onoma de M\'exico, AP 70543,
M\'exico DF 04510, Mexico.}

\begin{abstract}
The possible  imprint of quantum decoherence,  in the framework of cosmology, is here investigated. Particular attention is paid  to the observational fact that entanglement could lead to the  interaction of different eras of cosmic evolution. The role played by decoherence provides the existence of "quantum entanglement" between cosmological eras giving, as observational results,  dynamical constraints on the corresponding cosmological models. We analyze in detail the concept of linear entropy $S_L$ and of the so called Von Neumann entropy $S_{VN}$. Since these quantities are actually related to the measure of entanglement in physical systems, it is possible to quantify the degree of entanglement by using their definitions. The basic idea is to assume that the Von Neumann entropy is essentially the source of the cosmological entropy, in the simplest picture of a homogeneous and isotropic universe, with zero spatial curvature. Under this recipe, the definition of the cosmological constant could be re-obtained as an entanglement process between eras, if the cosmic states are built up by using different Hilbert spaces. In other words, an emergent cosmological constant naturally arises when the interaction of states produces a mixed density matrix, whose existence guarantees that the linear and the Von Neumann entropy exist. In addition, by investigating the dynamical properties of an evolving equation of state $\omega$, it is possible to show that a  dark energy model could be naively obtained, giving as a limiting result the $\Lambda$CDM model. In doing so, we focus on the so called negativity $\mathcal N$ between cosmological eras and on the dynamical effects of an evolving Von Neumann entropy on the standard thermodynamics of the universe. We recover the negativity functional form in terms of the scale factor $a(t)$ and  fix cosmological constraints, compatible with the convexity of $\mathcal N$. The functional form of $S$ and its evolution is also found to be bounded adopting the observable WMAP 7 years limits. The main result of such a procedure is to reproduce viable equations of state without introducing {\it a priori}  any vacuum energy  or cosmological constant. In this picture,  the cosmological constant can be seen as a first order Taylor expansion of a more general model, inferred from entanglement. This alleviates both the coincidence and the fine tuning problems. Numerical and observational bounds coming from the accelerated expansion of the universe  (dark energy) can be therefore reinterpreted  as the quantum signature of entanglement. Thus  by using the  cosmographic series related to the  luminosity distance,  it is possible to give  cosmological constraints on $S_{VN}$ and its derivatives. The physical meaning of these results  is twofold. On one hand, we fix constraints on $\omega$ and on the density of dark energy $\rho_{DE}$. On the other hand, it is possible to reinterpret  cosmographic parameters, i.e.  the acceleration $q$, the jerk $j$, the snap  $s$ and so forth, as variations of the Von Neumann entropy with respect to the redshift $z$. In conclusion, the quantum nature of entanglement allows a picture  of the  universe evolution where  dark energy effects are due to the existence of decoherence between two or more universe epochs.
\end{abstract}


\maketitle

\section{Introduction}

The challenge of unifying both quantum and classical aspects of gravity is an open question of modern  physics \cite{dtc1}. Unfortunately, a self consistent paradigm for  quantum gravity and a viable interpretation of quantum states in general relativity have not been clarified so far \cite{dtc2}. Quantum corrections seem to be important in order to alleviate the problems of standard cosmology, in which the simplest introduction of a vacuum energy cosmological constant results inadequate to characterize the whole universe dynamics \cite{adj}. Hence, quantum approaches in general relativity constitute a bid to formulate  (at least) a semi-classical scheme in which general relativity is replaced by more general paradigms. Additional hints of quantum effects in gravity emerge in the context of   unification schemes of fundamental interactions, such as superstring, supergravity, or grand unified theories, which generally exhibit effective actions containing non-minimal couplings to the geometry or higher order terms in the curvature invariants  \cite{sciama,sciama1}. For example, some possibilities have been considered, trying to investigate quantum field theory on curved spacetime, yielding corrections to the Einstein-Hilbert Lagrangian \cite{birrell}. It has been found that these corrective terms are however inescapable in the effective action of quantum gravity close to the Planck energy \cite{vilkovisky}. Clearly, all these approaches do not constitute the full quantum gravity theory, although they are needed as \emph{working frameworks} towards a self-consistent quantum scheme. On the other hand, the importance of measuring quantum gravity effects has led physicists to consider cosmological scales as the frameworks of quantum measurements at least in the infrared limit \cite{prl}.

In doing so, possible quantum signatures have been proposed  showing the possibility to detect measurable  effects  \cite{vile}. However, the thorny issue of understanding whether quantum effects are significative, in the  context of late time cosmology, remains an open debate \cite{ehb}. In particular, if quantum effects are significative in the observable universe, the nature of dark energy (DE) and dark matter (DM) could be inferred by a single approach where  quantum mechanisms should play an important role. A possible way  to provide the observed dynamics could be represented by the  entanglement between quantum states \cite{eccolo}. In particular, entanglement may represent the process between cosmic quantum states, which can drive the dynamics of the universe, as observed today. In other words, cosmological parameters, such as DE density, matter density, acceleration parameter, and so forth, may be reinterpreted as quantum signatures of cosmic quantum states, in the late time universe. However, one interesting point is wondering whether entanglement effects can affect the whole universe dynamics, not only at late times. To this regard, it is possible to   consider the so called decoherence of quantum states, where cosmological states are thought to be entangled to each other, through non-factorizable products of substates, associated to cosmological observables. Possible measurable effects can be detected by simply assuming a spatially flat universe. In particular, the cosmological constant  can be recovered as a limiting case of more general evolution. Each cosmological model, derived by assuming quantum decoherence of entangled cosmic states, depends on the choice of cosmic volume. We will show that once the cosmic volume is postulated, a corresponding barotropic factor naturally arises, showing repulsive effects, able to counterbalance the attraction of gravity. Moreover, a direct measurement of entanglement between cosmic quantum states can be directly related to the DE density, showing that it is possible to drive the cosmic speed up by the \emph{weight of entanglement} \cite{eccolo2}. The main purpose of this review is to highlight the basic demands of entanglement processes in cosmology, pointing out possible applications in the framework of observable cosmology, with particular attention  to the cosmography of the universe. We present an approach to build up a cosmic quantum state in terms of the  cosmographic parameters, finding extensions of the standard cosmological model.

The review is organized as follows: in Sec. II, we highlight the role played by entanglement, underlying the fact that it represents a resource of modern physics. We briefly describe its  possible role in  quantum gravity and quantum cosmology, with particular attention devoted to the concept of decoherence. In Sec. III, we deal with the observational problems of today  universe, emphasizing, in particular, the concept of DE. In Sec. IV, we develop a procedure for constructing entangled cosmological states, describing  the possible implications of the method. In particular, our technique allows  to show that the cosmological constant can be obtained from decoherence. In Sec. V, we stress the concept of decoherence, introducing the basic idea of entangled state ansatz. In Sec. VI, we introduce the role played by the so called Von Neumann entropy, as a source of varying DE terms. In Sec. VII, we investigate the cosmological implications of an entanglement measure, i.e. the so called negativity of entanglement. In particular, we show that modifications of  standard cosmological model occur, when the weight of entanglement becomes significative. Finally, Sec. VIII is devoted to conclusions and perspectives.

\section{Entanglement, Decoherence, and Quantum Cosmology}

Quantum information has reached much interest as entanglement became a resource of quantum physics \cite{eccolo3}. In particular, entanglement represents  a consolidate effect, whose application fields span from cryptography to teleportation and it is nowadays  strongly supported by several experiments. Entangled states can be assumed for different physical systems showing  the main feature of entanglement, i.e. the fact that no real information is carried out by a measure of entanglement.  In cosmology, the concept of  \emph{entangled quantum states} can be associated to each cosmological epoch. This turns out to be extremely relevant  since quantum systems can show the effects of entanglement  (and lead to possible quantum correlations) also at cosmological distances.

The main properties of entanglement rely on the fact that it origins from  complex structures and it is  sometimes difficult to be detected in  laboratory. However, the problem of determining the degree of entanglement  has reached much interest, providing nowadays a lot of entanglement \emph{rulers} that  quantify the weight of entanglement \cite{eccolo4}. Moreover,  mixed states should not be confused with entangled states. In fact, in the framework of entangled states, a physical system cannot be \emph{purified} from the property of being entangled, showing that entanglement is  an intrinsic  property of quantum states \cite{eccolo5}.

On the other side, even though the meaning of  entanglement for indistinguishable particles is still lacking, entanglement properties have been mainly investigated in the framework of special relativity. In particular, even in the relativistic case, the well known superposition principle is recovered and adopted in order to investigate quantum systems which do not behave as a mere product of substates. This is the basic demand of entanglement, i.e. it is not generally possible to assign a single vector to each subsystem. As a matter of fact, it is a standard convention to refer to the states satisfying such properties as entangled states, instead of separable states in the opposite case.

This feature could be extremely interesting for quantum gravity and quantum cosmology for the following reasons.

As we said, a unifying scheme dealing with quantum effects in general relativity represents an open question of modern physics  because no definitive paradigm is able to be predictive at interaction lengths of quantum gravity \cite{tze}. Furthermore,  the role played by gravity  at  laboratory scales is negligibly small so relevant gravitational effects are difficult to be detected in quantum field theory. Despite of this situation, it is a common opinion that effective quantum theories of gravity may represent additional steps towards the unification of all  physical interactions.

On the other hand, the standard cosmological model is based on assuming an initial \emph{naked singularity}, namely the \emph{Big Bang}, whose corresponding indirect effects seem to be detectable today. Since singularities may be interpreted as consequences of classical field theories, it would be more reasonable to infer a  quantum gravity approach which avoids  singularities at initial times. Conventional approaches are often grouped  in two main families. The first deals with covariant models which require the spacetime covariance, while the second ones, i.e. the canonical approaches, split the spacetime into a spatial and a temporal component, in analogy to quantum mechanics. In the last case, one chooses a foliation of spacetime, named $(3+1)$-decomposition, which split  spacetime into spatial hypersurfaces plus a temporal axis. In this framework, canonical coordinates and conjugate momenta are requested, in analogy to quantum theory. On the other side, quantum cosmology may represent the reduction of quantum gravity to simpler models, usually referred to as minisuperspaces. Thus, minisuperspaces can be seen as  cosmological models with a finite number of degrees of freedom, reduced from the  infinite numbers of quantum gravity degree of freedom.

Another interesting property is the fact that each quantum state, representing a given dynamical  system, could be thought as coupled with the environment. The way of describing the environment is to assume that it is represented by a large number of degrees of freedom. It turns out that a generic quantum state, which is entangled, can naturally be coupled with the environment. If one extends this picture to the environment only, one can notice that the environment couples with itself. As a consequence, the whole universe, described by a generic quantum system, i.e. by a particular wave function, can couple with all the macroscopic degrees of freedom. This process is known in  literature as \emph{decoherence}. The properties of decoherence naturally justify the assumption of cosmic quantum states, showing that the hypothesis of considering different entangled states for separate epochs of the universe can affect late time dynamics. This is the main hypothesis of this paper that we are going to discuss below.

\section{Late time Cosmology and the problem of Dark Energy}

 The discovery of the observed cosmic speed up \cite{SNeIa1,SNeIa2,CMBR1,CMBR2,LSS1,LSS2,Lyalpha} showed that general relativity  is inadequate to describe  \emph{in toto} the universe dynamics unless huge amounts of unknown cosmological fluids are  assumed. In fact, the presence of baryons and cold DM predicts a decelerating universe, in contradiction with current observations \cite{vnt1,vnt1bis}. Moreover, the advent of the so called {\it precision cosmology} \cite{wei} has been capable of probing physics at higher redshift with increasing precision, fixing tighter constraints on the observable universe. As a natural consequence, a new ingredient \cite{vnt2,vnt2bis} is required in the cosmological puzzle \cite{vnt3,vnt3bis}, showing a negative pressure able to reproduce the late time positive acceleration. In particular, recent observations definitively forecasted that about 70$\%$ of the universe is filled with this unknown ingredient, usually treated as an exotic fluid, counterbalancing the attractive effect of gravity. The challenge of discovering the nature of this fluid represents an open issue of modern astrophysics. The corresponding cosmological component, which is thought to drive the observed acceleration, is dubbed DE \cite{pad}, and seems to behave as a non-interacting perfect fluid, whose equation of state (EoS), i.e. $\omega$, is bounded by present observations around $\omega \simeq -1$. As a first possible explanation, a vacuum energy cosmological constant, namely $\Lambda$, has to be  taken into account in order to characterize this exotic fluid  in its simplest, non-evolving form \cite{Lambda}. However, the corresponding model, namely $\Lambda$CDM, suffers from two thorny drawbacks, i.e. the so called fine-tuning and coincidence problems \cite{wei2} and consequently it is believed that the $\Lambda$CDM model cannot be seen as the final cosmological paradigm, being rather a limiting case of a more general theory. 
 
 Thence, several alternatives have been proposed, spanning from phenomenological DE terms to modifications of the Einstein-Hilbert action. An immediate and natural extension of $\Lambda$CDM is the so called $\omega$CDM model \cite{coppa}, also referred to as Quintessence model, described by introducing a scalar field $\phi$ \cite{vnt3}, which guarantees the existence of a constant EoS \cite{blundo}. Analogously to $\Lambda$CDM, the origin and the physical meaning of the involved scalar field has not been  clarified, and so the $\omega$CDM model cannot be considered as a final model. Thus, in order to give an explanation which takes into account  an evolving DE, varying Quintessence models have been accounted  \cite{sa}. This approach  leads to the introduction of time dependent barotropic factor, i.e. $\omega=\omega(z)$, derived  by the ratio between the pressure and density, i.e. $\omega(z)=\frac{\mathcal{P}}{\rho}$. The main disadvantage of assuming an evolving barotropic fluid is that the form of $\omega(z)$ is usually not defined \emph{a priori}, and may  be reconstructed and bounded by observations \cite{ai1,ai2,ai3}.

In this approach, barotropic fluids allow to preserve causality. This turns out to be a consequence of assuming a positive (adiabatic) speed of sound, i.e. $c_s>0$. Moreover, considering the first derivative of $\omega$ with respect to $\ln a$, i.e.
\begin{equation}
\omega' = \frac{d\omega}{d\ln a} \,,
\end{equation}
related, in models involving a scalar field, to the slow-roll potential region $z = 1$, one gets for barotropic fluids
\begin{equation}
\label{Barotropic}
\omega' < 3\omega(1+\omega)\,.
\end{equation}
A possible example is the so called  Chevallier, Polarsky, Linder (CPL) parametrization of the EoS \cite{cpl}. In this picture, one expands in series the barotropic factor at the first order in $a(t)$ series, i.e. $\omega(a)=\omega_0+\omega_1(1-a)$. Hence, by assuming $a\equiv(1+z)^{-1}$, one finds ${\displaystyle \omega(z)= \omega_0 + \omega_1 \frac{z}{1+z}}$, which provides, at low and at high redshifts, $\omega(z\rightarrow0)=\omega_0$ and  $\omega(z\rightarrow\infty)=\omega_0+\omega_1$, respectively.

Other classes of DE models, similar to barotropic fluids, are  ``thawing'' models, where
\begin{equation}
1+\omega \lesssim \omega' \lesssim 3(1+\omega)\,,
\end{equation}
or "cooling" models, where
\begin{equation}
\label{Cooling}
-3(1-\omega)(1+\omega) < \omega' \lesssim f(z) \, \omega(1+\omega) \,,
\end{equation}
where $f(z)$ is a parametric function of the redshift. Finally, the so called phantom models are represented by
\begin{equation}
\label{Phantom}
\omega < -1 \,.
\end{equation}

As reported in Sec. I, another typical approach is to extend general relativity by modifying gravity through curvature DE or quantum gravity modifications \cite{sciama1, recapo}. Such models provide additional terms to the standard Ricci scalar, showing that DE effects can be recast in a curvature fluid. Without enter into  details, the need of comparing cosmological models with data, through the use of standard distance indicators, increased its importance in order to alleviate the  degeneracy problem between cosmological models. It is a matter of fact that different models could fit the same cosmological data with high precision. In other words, a single  test performed by standard candles, usually is affected by a strong model dependence. The reasons rely on the fact that the model under exam is assumed to be the best one, therefore numerical tests suffer of a thorny degeneracy in fitting cosmological data. The standard observational tests, i.e. by Supernovae Ia, baryonic acoustic oscillations and cosmic microwave background are usually combined between them in order to reduce the phase space region in which the best fits are confined. A superposition of different cosmological tests may alleviate the degeneracy problem. Several parameterizations of the EoS of DE correspond to different classes of DE models, providing different regions of allowed cosmological observables. In this review, we discuss the cosmological models derived from the entanglement between cosmic quantum states, showing that it is possible to infer a barotropic EoS directly related to entanglement and decoherence.

\section{Cosmological entangled states}

It is a matter of fact that entanglement between observables in the universe may emerge in the framework of quantum gravity, through the process of decoherence. However, the interpretation of this point  turns out to be very  intricate    since a self consistent  theory of quantum gravity is still lacking. A first example comes from   the solutions of the  Wheeler-De Witt  \cite{Hartle:1986gn} which may be considered  as  examples of entangled quantities. On the other hand,  dynamical effects like DE and DM could be reduced to in a unitary paradigm using entangled cosmological states \cite{eccolo}. In this picture one interprets the hidden mechanisms behind DE and DM as a first evidence of quantum  signatures. In particular, DE can be treated in terms of a cosmological constant, i.e. $\Lambda$, whose value is today constrained with increasing accuracy, showing a magnitude comparable to standard pressureless matter. This property reveals several implications for observations \cite{tuturru}. Thence, by looking at the universe as a whole system, characterized by a quantum wave function, one may interprete the "classical" cosmological constant in terms of quantum observables.

The underlying philosophy takes into account the entanglement of macroscopic degrees of freedom, leading to classical \emph{appearances} of macroscopic systems (in this case, the whole observed universe), through the process of decoherence \cite{deco}. Under these hypotheses, a possible connection between quantum information and quantum cosmology can be easily postulated. In other words, quantum cosmology may emerge since microscopic degrees of freedom, characterizing earlier phases of the universe dynamics, are thought to be entangled to each other. For now on, the word \emph{state} represents the dynamical properties at a certain time of the universe evolution.


 The reasons of considering entanglement between cosmological states is twofold. First, the achievement that entanglement phenomena between quantum states have been framed in robust theoretical schemes and verified through several experimental tests (see for example \cite{guehne,elk,carteret,horodecki,horodecki2,112,walborn,harald,vidal,bell,chsh}). Thence, at earlier phases of the universe dynamics, it is possible that quantum wave functions may be entangled, showing complex behaviors and providing the observed universe today. 
 Second, casual regions of the universe are related through inflationary phases, by adopting scalar fields, whose introduction is performed by hand. If entanglement between quantum states occurred, casual regions can be naturally related to each other in a simple way.

Hence, assuming the validity of the superposition principle and the non-separability of the quantum states, a compound cosmological quantum system cannot be factorized into
pure local states, leading to a \emph{non-factorizing property} \cite{libroentanglato}, which can be seen as responsible of casual connections between states, without invoking an inflationary phase. A generic quantum cosmological state has the form
\begin{equation}\label{statusaaa}
|\Psi\rangle\propto|\Psi^{(1)}\rangle|\Psi^{(2)}\rangle\,,
\end{equation}
well-factorized into two sub-states, i.e. $|\phi_1\rangle$ and $|\phi_1\rangle$, corresponding to
\begin{eqnarray}\label{kj09864}
|\phi_1\rangle= |\Psi^{(1)}\rangle\,,\nonumber\\
\,\\
|\phi_2\rangle= |\Psi^{(2)}\rangle\,,\nonumber
\end{eqnarray}
which indicate that Eq. ($\ref{statusaaa}$) is a separable state. On the other hand, a state of the form
\begin{equation}\label{statusaaa2}
|\Psi\rangle\propto\sum_{n=0}^{M}|\Psi^{(1)}_{n}\rangle|\Psi^{(2)}_{n}\rangle\,,
\end{equation}
provides more information, since it leads to a complex physical system, built up by substates corresponding to different epochs of the universe. It turns out that the information of entanglement is here emphasized, without the possibility to factorize in substates the wave function of Eq. ($\ref{statusaaa2}$). A possible interpretation of scalar fields enters the definition of quantum states. In fact, by assuming
\begin{equation}\label{kjhdwkhfdj}
\phi_n(t)=\langle x|\phi_n\rangle\,,
\end{equation}
it is possible to interpreter the existence of scalar field, i.e. $\phi(t)$, in cosmology, by simply assuming that $\phi(t)$ represents the projector of the n$-th$ $|\phi_n\rangle$ state on the set $\langle x|$, whose elements are represented by quantum cosmological observables, in the case $n=1$. One of the most relevant properties of assuming Eq. ($\ref{statusaaa2}$) for characterizing the universe dynamics, instead of taking into account Eq. ($\ref{kj09864}$), is that a viable technique to investigate the dynamics may be performed in terms of the  {\it density matrix}, i.e. $\hat\rho$.

In the framework of $\hat \rho$, one \emph{prepares} the state $|\Psi_1\rangle$ by considering a probability $p_1$, the state $|\Psi_2\rangle$ with probability $p_2$ and so forth, for all the sub-states of the whole system, i.e. the universe. This procedure allows  to find out a quantum signature between states, even if two systems are spatially far \cite{73}. This is a consequence of the fact that for a bipartite (or multipartite) quantum state \cite{111}, the non-separability (or non-factorizability) provides correlations, which evolve in terms of the cosmic redshift. One can wonder whether the so called entanglement-degree  can quantify how much a quantum state is mixed or pure \cite{112,13,101,177}, i.e. the so called \emph{weight} of entanglement, that we cited in the above sections. Meanwhile, the need of assuming the validity of standard thermodynamics may show the emergence of the cosmological constant $\Lambda$, as a result of quantum entanglement. In doing so, let us consider a $N$-dimensional Hilbert space, where the probabilities are given by
\begin{equation}\label{getosjag}
p_k =
\frac{1}{N}\,,
\end{equation}
valid $\forall k$ and associated to our ignorance about the configuration of a certain quantum cosmological state. In fact, if one of the $p_k$ is different from zero, we find a pure quantum state, entangled otherwise. For a generic pure state $|\Psi\rangle$, the density matrix is the projector
\begin{equation}\label{matriosca}
\hat\rho_p=|\Psi\rangle\langle\Psi|\,.
\end{equation}
The main properties of $\hat \rho$, which may be satisfied in the framework of the universe, are:
\begin{equation}
Tr\hat\rho=1,
\end{equation}
corresponding to sum over the whole set of eigenvalues of $\hat \rho$;
\begin{equation}\label{dln222222222}
Tr\hat\rho^2\leq1,
\end{equation}
and
\begin{equation}\label{nau3iuy7387y}
\langle\chi|\hat\rho|\chi\rangle\geq 0,
\end{equation}
representing a convex sum over $\hat\rho^2$, i.e.  Eq. ($\ref{dln222222222}$), whose mean value, over a given state $|\chi\rangle$, is positive defined, i.e. Eq. ($\ref{nau3iuy7387y}$) holds.


 It is important to remark that the above properties lead to
\begin{equation}\label{eosp}
Tr\hat\rho^2=1\,,
\end{equation}
only if the state is separable, otherwise the condition $Tr\hat\rho^2<1$ holds, providing an indication of entanglement between quantum states. In other words, the condition $Tr\hat\rho^2<1$ is a first prototype of entanglement measurement. The need of more accurate measures of entanglement is however extremely important to quantify the weight of entanglement. In particular, we define the \emph{purity} of a state as
\begin{equation}\label{mnmnnnbnbnn}
\mu\left[\hat\rho\right]=Tr\hat\rho^2\,,
\end{equation}
and then  the basic concept of \emph{linear} \emph{entropy}
\begin{equation}\label{bnbnbnbnbnbn}
S_L=\frac{N}{N-1}\left(1-\mu\left[\hat\rho\right]\right),
\end{equation}
which are two prototypes of entanglement measurements. For now on, we are interested in quantifying a more general definition of entanglement entropy, since we need to correlate the corresponding effects of entanglement to the cosmological entropy.

 The validity of standard thermodynamics is clearly a request of observational cosmology for late time universe. In particular, the energy density, $\rho$, of our universe is a component of the energy momentum tensor, i.e.
\begin{equation}\label{tenxxxx}
T^{\mu\nu}=(\rho +\mathcal{P})u^{\mu} u^{\nu} - \mathcal{P} g^{\mu \nu}\,,
\end{equation}
written in the perfect fluid form  and satisfying the conservation law
\begin{equation}\label{d45ndkjn}
\nabla^\mu T_{\mu\nu}=0\,,
\end{equation}
which gives the continuity equation
\begin{equation}\label{lej2}
\frac{d\rho}{dz}=3\frac{ \mathcal{P} +\rho}{1+z}\,,
\end{equation}
with the definition of standard cosmic time, in terms of the redshift  $z$, i.e.
\begin{equation}\label{vvvvvvnnnnnnnnn}
\frac{dz}{dt}=-(1+z)H(z)\,,
\end{equation}
where $H$ is the Hubble parameter. In this picture, contributions come from DE into Eq. ($\ref{tenxxxx}$) are usually added  by hand. However, if one assumes that $\rho=\sum_i\rho_i$, where $\rho_i$ represents the single energy budget of each species, it is licit  to assume that
\begin{equation}\label{energiaentropia}
\frac{\partial S}{\partial \rho}=\frac{1}{T}\,,
\end{equation}
where, for our quantum states,  the whole entropy $S$ is the so-called {\emph Von Neumann entropy}, i.e. $S\sim S_{VN}$, which usually describes how large is the lack of information of a quantum state in terms of the temperature $T$, i.e.
\begin{equation} \label{temp}
\mathcal{T} \propto  \exp\left[{3\int{\frac{ \omega(z)}{1+z}}dz}\right]\,.
\end{equation}
Additional contributions to $S_{VN}$ are given by standard constant entropy density given by pressureless matter, which is however not able to drive the today universe  towards  a positive accelerated state. Since thermal entanglement properties of the universe are crucial in the definition of the net energy budget, we need a measure of the entropy of the whole system, i.e. the universe. The so called Shannon-Von Neumann entropy reads
\begin{equation}\label{enggg}
\hat S=-Tr\left(\hat{\rho}\ln\hat{\rho}\right)\,=\,-\sum_k \lambda_k\ln
\lambda_k\,,
\end{equation}
which satisfies concavity, subadditivity and Araki-Lieb inequality \cite{orlando}.

Since we are assuming an arbitrary choice for quantum cosmological states, the corresponding eigenvalues of $\hat \rho$, i.e. $\lambda_k$, can define the whole set of cosmological observables, by construction. This turns out to be compatible with the fact that the entropy $S$ of a given system represents a macroscopic manifestation of microscopic constituents. It follows that, measuring entanglement between quantum states of the universe would represent a "coarse grained" approach capable of contributing to the global description of the universe. Our purpose is to demonstrate that the cosmological constant density, i.e. $\Omega_\Lambda$, may be interpreted as a result of entangled cosmic quantum states.

\section{The entangled state ansatz}

Now we are interested in describing the above states in terms of observables by quantifying the effects due to entanglement. Hence, let us take into account a set of observable quantities which defines a classical cosmological state, written in terms of the following vector
\begin{eqnarray}\label{setto}
\Upsilon\,=\,\left\{H(z), a(z), q(z), j(z), \Omega_{m}(z), \Omega_{k}(z), \Omega_{\Lambda}(z),T(z),\ldots\right\},
\end{eqnarray}
where  $q(z),j(z)$ are the  cosmographic parameters, respectively the
{\it acceleration} parameter and the variation of acceleration, namely the {\it jerk} parameter.  $\Omega_{m}$ is the normalized matter
density, while $\Omega_{k}$ is the normalized spatial curvature density. All the parameters are assumed as function of the redshift $z$ and univocally assign a cosmological state. In particular, the cosmographic parameters can be inferred by expanding the scale factor into a series around a given time $t_0$,
yielding
\begin{eqnarray}\label{eq:expa}
a(t)  =  a_0\cdot \Bigl[ 1 + \frac{da}{dt}\Big|_{t_0} (t-t_0) +\,\frac{d^2a}{dt^2}\Big|_{t_0} (t-t_0)^2 +\frac{d^3a}{dt^3}\Big|_{t_0} (t-t_0)^3+\ldots\Bigr]\,,
\end{eqnarray}
where we truncated the series at the third order in $\Delta t\equiv
t-t_0$. Since $t-t_0>0$ and conventionally $a_0=1$, we recast Eq.~($\ref{eq:expa}$) as
\begin{eqnarray}\label{serie1a}
a(t)  =   1 - H_0 \Delta t - \frac{q_0}{2} H_0^2 \Delta t^2 - \frac{j_0}{6} H_0^3 \Delta t^3 +\ldots\,.
\end{eqnarray}
Here, the subscript $"0"$ in Eq. ($\ref{serie1a}$) indicates that the coefficients are evaluated at
$t=t_0$. In the literature, such coefficients are also named the cosmographic set \cite{cosmografiaorlando}. The physical meaning of $q$ specifies whether the universe is accelerating or not, depending on its sign. An accelerating universe requires
$-1\leq q_0<0$, while a positive $j_0$ implies that $q$ changes sign as the universe expands. For these reasons, $j$ is also known as the variation of acceleration. Today it is known with accuracy of $4.2\sigma$ that the universe is positively expanding, so the quest of determining the sign of $j$ is  an open challenge for determining at which redshift the acceleration starts. The most important advantage associate to Eq. ($\ref{serie1a}$) is that the definitions of $q$ and $j$ does not depend on the choice of a particular cosmological model. In other words, expanding $a(t)$ in powers of $\Delta t$, does not depend on cosmology. However, by the definition of the Hubble rate, one can relate the cosmographic parameters to $H$, yielding
\begin{eqnarray}\label{eq:CSoftime}
q&=&-\frac{\dot{H}}{H^2} -1\,, \nonumber\\
\,\\
j&=&\frac{\ddot{H}}{H^3}-3q-2\,, \nonumber
\end{eqnarray}
so that, by following the procedure proposed in \cite{cosmografiaorlando}, it is possible to measure $q$ and $j$ at our time independently from the choice of $H$. Keeping these results in mind, we assume a minisuperspace where the cosmological principle holds and "evolution" is replaced by correlation among states. In some sense, we are defining  a cosmography for any given era, assuming a correlation without specifying a priori the cosmological dynamical equations.

In the vector $\Upsilon$, the cosmographic parameters capable of describing the universe dynamics are therefore taken into account. A possible \emph{minimal choice} of observable quantities, which  can be used to evaluate a complete phase-space evolution, through the use of entanglement, i.e. the minimal set of parameters which may characterize the universe evolution, provides the introduction of $\Omega_k$ and $\Omega_m$, once $H_0$ is fixed. The others,
as for example the remaining cosmographic parameters, can be expressed in terms of $\Omega_k$ and $\Omega_m$. This  minimal choice  allows to write down
\begin{equation}\label{stati}
|\phi_{i}\rangle\equiv \left(\begin{array}{c}
\Omega_{mi}\\
\Omega_{ki}
\end{array}
\right),
\end{equation}
where we labelled $i=1,2,\ldots,N$ for indicating a generic era.

For the sake of simplicity, let us take into account only two regions, referring to the \emph{Entangled States Ansatz} (ESA) as the basic hypothesis to construct a superposition of states. In terms of a basis, following the Gram-Schmidt construction of orthogonal and normalized basis, we have
\begin{equation}\label{stati1ver}
|e_{1}\rangle= N_{1}\left(\begin{array}{c}
\Omega_{m1}\\
\Omega_{k1}
\end{array}
\right),
\end{equation}
which is the first unitary vector;  the other one is built up by considering the rule,
\begin{equation}\label{stati2ver}
|\tilde{e}_{2}\rangle=\left(\begin{array}{c}
\Omega_{m2}\\
\Omega_{k2}
\end{array}
\right)-N_{1}^{2} \left(\begin{array}{cc} \Omega_{m1} & \Omega_{k1}
\end{array}\right)\cdot
\left(\begin{array}{c}
\Omega_{m2}\\
\Omega_{k2}
\end{array}
\right) \left(\begin{array}{c}
\Omega_{m1}\\
\Omega_{k1}
\end{array}
\right),
\end{equation}
which can be normalized as
\begin{displaymath}
|e_2\rangle=N_2\left(
\begin{array}{c}
\Omega_{m2}-N_{1}^{2}\left(\Omega_{m1}^{2}\Omega_{m2}+\Omega_{m1}\Omega_{k1}\Omega_{k2}\right)\\
\\
\Omega_{k2}-N_{1}^{2}\left(\Omega_{k1}^{2}\Omega_{k2}+\Omega_{m1}\Omega_{k1}\Omega_{m2}\right)
\end{array}
\right),
\end{displaymath}
where the normalization parameters $N_1$ and $N_2$ are given by
\begin{widetext}
\begin{eqnarray}
N_1\,\equiv\,\frac{1}{\sqrt{\Omega_{m1}^{2}+\Omega_{k1}^{2}}},\,\,\,\,\,\,\,\,\,\,\,\,\,\,\,\,\,\,\,\,\,\,\,\,\,\,\,\,\,\,\,\,\,\,\,\,\,\,\,\,\,\,\,\,\,\,\,\,\,\,\,\,\,\,\,\,\,\,\,\,\,\,\,\,\,\,\,\,\,\,\,\,\,\,\,\,\,\,\,\,\,\,\,\,\,\,\,\,\,\,\,\,\,\,\,\,\,\,\,\,\,\,\,\,\,\,\,\,\,\,\,\,\,\,\,\,\,\,\,\,\,\,\,\,\,\,\,\,\,\,\,\,\,\,\,\,\,\,\,\,\,\,\,\,\,\,\,\,\,\,\,\,\,\,\,\,\,\,\,\,\,\,\,\,\,\,\,\,\,\,\,\,\,\,\,\,\,\,\,
\\
N_2\,\equiv\,\frac{1}{\sqrt{\left\{\Omega_{m2}-N_{1}^{2}\left(\Omega_{m1}^{2}\Omega_{m2}+\Omega_{m1}\Omega_{k1}\Omega_{k2}\right)\right\}^{2}+\left\{\Omega_{k2}-N_{1}^{2}\left(\Omega_{k1}^{2}\Omega_{k2}+\Omega_{m1}\Omega_{k1}\Omega_{m2}\right)\right\}^{2}}}.
\end{eqnarray}
\end{widetext}
From the above construction, we may postulate the following cosmological state
\begin{equation}\label{signdivin}
|\Psi_{\pm}\rangle\,=\,\alpha|e_1\,e_1\rangle\pm\beta|e_2e_2\rangle,
\end{equation}
with the direct expression for $|e_i e_i\rangle$
\begin{eqnarray}\label{mkooo}
|e_i\,e_i\rangle=N_{i}^{2}\left(\begin{array}{c}
\Omega_{mi}^{2}\\
\Omega_{mi}\Omega_{ki}\\
\Omega_{mi}\Omega_{ki}\\
\Omega_{ki}^{2}
\end{array}
\right),
\end{eqnarray}
and the normalization given by
\begin{equation}\label{normalyyy}
|\alpha|^2+|\beta|^2=1\,.
\end{equation}
If we define the two {\emph simplest positions} for matter and curvature
\begin{equation}\label{ajla}
\Omega_{m2}^{*}=\Omega_{m2}-N_{1}^{2}\left(\Omega_{m1}^{2}\Omega_{m2}+\Omega_{m1}\Omega_{k1}\Omega_{k2}\right)\rightarrow\Omega_{m2}\,,
\end{equation}
and
\begin{equation}\label{djfdj88262}
\Omega_{k2}^{*}=\Omega_{k2}-N_{1}^{2}\left(\Omega_{k1}^{2}\Omega_{k2}+\Omega_{m1}\Omega_{k1}\Omega_{m2}\right)\rightarrow\Omega_{k2}\,,
\end{equation}
we can recast $\Omega_{m2}$ and $\Omega_{k2}$ as functions of $\Omega_{m1}$, $\Omega_{k1}$. A maximally entanglement phenomenon happens when 
$\alpha=\beta=\frac{1}{\sqrt{2}}$, up to a phase factor.

Finally, recasting the eigenvalues of the density matrix, one gets for the first era as 
\begin{equation}\label{firstera}
  \Omega_{m1}+\Omega_{k1}+\sum_{i}\Omega_{X1i}=1,
\end{equation}
and analogously,  the second one
\begin{equation}\label{secondera}
  \Omega_{m2}+\Omega_{k2}+\sum_{j}\Omega_{X2j}=1.
\end{equation}
Immediately, the interpretation of $\Lambda$ arises  since it is possible to connect its existence with the lack of information of an entangled
state, described by the Von Neumann entropy. In other words, one can interpreter the additional term, i.e. $\Omega_{Xh}$ as the result of the entanglement between two quantum states.


This procedure can be immediately generalized.  Even though a description of the whole universe can be found in terms of entangled states, it is remarkable to notice that each epoch can be defined by its own reduced density matrix. In other words, each substate can be imagined as entangled, providing a reduced density matrix, defined as
\begin{equation}\label{defmatrxred}
\hat \rho^{A}=Tr_{B}\hat \rho,
\end{equation}
where we assumed $\hat\rho=\hat\rho^{AB}$, for two different spaces, $A$ and $B$, corresponding to sub-parts of a given era. This permits us to define a cosmological constant constrained to our time, i.e. $z=0$.

By considering the eigenvalues of $\hat\rho$, i.e.
\begin{equation}\label{k}
S\,=\,-\left(\lambda_3\ln\lambda_3+\lambda_4\ln\lambda_4\right),
\end{equation}
in the case of
$\Omega_{k1}\approx 0$, $p_{+}=p_{-}$ and for maximally entangled states, we infer the expression for
$\Omega_{\Lambda}$, in the case of
$\Omega_{k1}\approx0$
\begin{widetext}
\begin{equation}\label{OMEGALAMBDA22}
\Omega_{\Lambda}=\frac{\left(\Omega_{m2}^{2}+\Omega_{k2}^{2}\right)\Omega_{m1}^{5}-\left(p_-\Omega_{m2}^{2}-\Omega_{m2}^{2}-\Omega_{k2}^{2}/2\right)\Omega_{m1}^{4}-2\Omega_{m1}^{3}\Omega_{m2}^{2}+2p_+\Omega_{m1}^{2}\Omega_{m2}^{2}+\Omega_{m1}\Omega_{m2}^{2}+2p_-\Omega_{m2}^{2}-\Omega_{m2}^{2}}{\left(2\Omega_{m1}^{2}-1\right)\Omega_{m2}^{2}-\left(\Omega_{m1}^{2}+\Omega_{k1}^{2}\right)\Omega_{m1}^{4}},
\end{equation}
\end{widetext}
which confirms the fact that it is possible to \emph{directly infer} the value of the cosmological constant through the ESA at our time.

In general, the most general expression of $\Omega_\Lambda$ is a complicated function of $\Omega_{m,1}$, $\Omega_{m,2}$, $\Omega_{k,1}$ and $\Omega_{k,2}$.  Moreover, accordingly to our picture, a more general condition can be performed by assuming $N$ states, characterized by the vector set, as reported above. Independently from the grade of entanglement, the cosmological constant can be seen as a direct consequence of ESA. On the other hand, the coincidence problem is accounted by the fact that, by construction, $0<\Omega_{\Lambda}<1$, providing the observational range of $\Omega_\Lambda$, whose magnitude is naturally comparable with $\Omega_m$. This result naturally can solve the so called coincidence problem of cosmological constant.

\section{The Von Neumann entropy as a source for running Dark Energy}

The basic postulate of assuming entanglement, as peculiar mechanism which relates different cosmological eras, is intimately connected to the  degrees of freedom  of macroscopic systems. In particular, we consider the  whole universe in terms of entangled wave functions, by using the above  mechanism of decoherence, where the interaction between quantum states and  environment is assumed. This turns out to match quantum coherence and entanglement of cosmological states, leading to decoherence as soon as  the number of qubits increases. Hence, one can wonder whether quantum decoherence is able to leave an imprinting in the context of observational cosmology. Signatures of decoherence can be found at any epoch, by tracing the DE evolution when the Von Neumann entropy evolves as thermodynamic entropy. This property allows  to imagine an evolving DE term, derived from the Von Neumann entropy, which becomes a source of DE, at any epochs. As a consequence, the existence of DE can  be seen as a quantum signature derived from decoherence processes, as the volume evolves. We investigate two cases:
\begin{equation}\label{nhytg}
V\propto a^3\,,
\end{equation}
which corresponds to an evolving adiabatic volume, whose physical properties have been extensively described in \cite{mioagain} and the volume
\begin{equation}\label{kmoiutf}
V\propto H^{-3}\,,
\end{equation}
which depends on the apparent radius $R_A$ as described in \cite{losotros}. It is important to remark that both choices deal with a functional volume of the form 
\begin{equation}\label{ffgfg}
 V=V_0R^{3}\,,
\end{equation}
where $R$ is a suitable  {\it universe radius}, whose properties refer to as $V\rightarrow0$ as $z\rightarrow\infty$ and ${\displaystyle V_0\equiv\frac{1}{H_0^3}}$ when $z=0$. More complicated volumes have not treated in this review.

The corresponding cosmological models extend the $\Lambda$CDM paradigm, which becomes a limiting case of a more general theory. Under these hypotheses,  entanglement may represent a source for DE. Thus, relating the Von Nuemann entropy to standard entropy as 
\begin{equation}\label{kjefwcoss}
\frac{dS_{VN}}{dz}\Big|_{z>0}=\frac{dS}{dz}\Big|_{z>0}\,,
\end{equation}
for all $z$, we can write down
\begin{equation}\label{rhox}
-\frac{d\rho}{dz}\left(\log\rho+1\right)=\frac{1}{T}\frac{d\Big[\left(P+\rho\right)V\Big]}{dz}\,,
\end{equation}
where we made use of the continuity equation for a given volume $V$. Particularly, for the two volumes, we find
\begin{eqnarray}\label{tqjois}
3\frac{\log(e\rho)}{a^{3}}+\frac{1}{T}\Big[3 \omega(a)+\frac{1}{a}\frac{d\log(1+\omega(a))}{dz}\Big]=0\,,
\end{eqnarray}
and
\begin{eqnarray}\label{tqjois2}
3\frac{\rho}{a}\log(e\rho)(1+\omega(a))-\frac{1}{a\rho^2 T}\Big[6 (1+ \omega(a))^2-\frac{1}{a}\frac{d\omega}{dz}\Big]=0\,,\,\,\,
\end{eqnarray}
respectively for $V\propto a^{3}$ and $V\propto H^{-3}$ where  $e$ is the Napier number.
The cosmological constant can be recovered at high redshift, for both the two models, while at small redshift small corrections occur, i.e.

\begin{equation}\label{w1}
\omega^{(1)}\sim \alpha + \beta \frac{1}{\beta + \gamma(1+z)^n}\,,
\end{equation}
and
\begin{equation}\label{w2}
\left\{
  \begin{array}{ll}
    \omega^{(2)}\sim -\delta^2 z, & \hbox{$0 \leq z\leq 1$;} \\
    \omega^{(2)}\sim \epsilon^2 z, & \hbox{$z\geq 1$,}
  \end{array}
\right.
\end{equation}
respectively for the two models. Here $\alpha, \beta, \gamma, \delta, \epsilon$ are free parameters, matched by observations.
In particular, it seems that the introduction of entanglement processes, evaluated as a consequence of decoherence in the observable universe, naturally reproduces phenomenological equations of state, without assuming  any cosmological constant \emph{a priori}. The corresponding barotropic factors are similar to well known  parameterizations of the EoS of DE. An important result is that we do not put by hand any corrections to $\omega$ a priori, showing that it is possible to recover the behavior of DE in terms of entanglement only \cite{orlando}.

\section{Extending $\Lambda$CDM through entanglement measures: The case of negativity}

Another possibility to characterize the late time dynamics of the universe is offered by considering different measures of entanglement. We investigated the effects of the Von Neumann entropy, showing that cosmological models can be inferred once the volume evolves with $z$. On the other hand, assuming  the simplest multipartite system for two epochs, characterized by the vectors
\begin{eqnarray}
 |e_A\rangle_{1} |e_A\rangle_{2},\;
  |e_A\rangle_{1} |e_B\rangle_{2},\;
   |e_B\rangle_{1} |e_A\rangle_{2},\;
    |e_B\rangle_{1} |e_B\rangle_{2},
 \label{basis}
\end{eqnarray}
in a corresponding  Hilbert space $\mathbb{C}^2\otimes \mathbb{C}^2$, we may rewrite the more general state as
\begin{equation}
|\Psi\rangle = \alpha|e_A\rangle_1|e_B\rangle_2 + \beta|e_B\rangle_1|e_A\rangle_2\,,
\label{stato}
\end{equation}
considering the so called negativity of entanglement as a more complicated measure of entanglement \cite{orlando2}.

We define the negativity as
\begin{equation}\label{ns}
\mathcal{N}=2\sum_k \max(0,-\lambda_k)\,,
\end{equation}
where the sum is over the eigenvalues of the partially transposed density matrix \cite{altro1000}. Our choice is to parameterize the negativity as follows
\begin{equation}\label{nocomment}
\mathcal{N}(z)= N_0+N_1(1+z)^2\,,
\end{equation}
where we expanded $\mathcal{N}$ in power series  of $a^{-1}$. Terms $\propto a^{-1}$ have not been considered in agreement with phenomenological parameterizations  of the EoS of DE \cite{fine1}. The density matrix $|\Psi\rangle\langle\Psi|$ is
\begin{eqnarray}
\left(
\begin{array}{cccc}
0 & 0 & 0 & 0\\
0 & |\alpha|^2 & \alpha\beta^* & 0\\
0 & \alpha^*\beta & |\beta|^2 & 0\\
0 & 0 & 0 & 0
\end{array}
\right)\,,
\end{eqnarray}
and the corresponding partial transpose reads
\begin{eqnarray}
\left(
\begin{array}{cccc}
0 & 0 & 0 & \alpha\beta^*\\
0 & |\alpha|^2 & 0 & 0\\
0 & 0 & |\beta|^2 & 0\\
\alpha^*\beta & 0 & 0 & 0
\end{array}
\right),
\end{eqnarray}
giving
\begin{equation}
\mathcal{N}=2|\alpha\beta|\,,
\label{negpsi}
\end{equation}
where we made use of the eigenvalues, i.e.
\begin{equation}
  \lambda_1 = |\alpha|^2,\;
  \lambda_2 = |\beta|^2,\;
  \lambda_3 = |\alpha\beta|,\;
  \lambda_4 = -|\alpha\beta|.
  \label{evalT1}
\end{equation}
Once the eigenvalues have been found, we need to relate $\alpha$ and $\beta$ to cosmological densities. It is easy to get
\begin{equation}\label{modal}
|\alpha|=\frac{N_0+N_1(1+z)^2}{2|\beta|}\,,
\end{equation}
and
\begin{equation}
\Omega_m+\Omega_k+\Omega_X=|\alpha|^2+|\beta|^2\,.
\label{1}
\end{equation}
By approximating at low redshift our solution, we get
\begin{equation}\label{OmX}
\Omega_X(z)=|\beta|^2+\frac{\left(N_0+N_1(1+z)^2\right)^2}{4|\beta|^2}
- \Omega_m(1+z)^3\,,
\end{equation}
which represents the DE density in terms of the negativity, i.e. parameters $N_0, N_1$.
The Hubble rate is
\begin{equation}
H(z)=H_0\left[|\beta|^2+\frac{\left(N_0+N_1(1+z)^2\right)^2}{4|\beta|^2}\right]^{1/2}\,,
\label{Hubble}
\end{equation}
which drives the dynamical evolution of the universe in terms of the Friedmann equations, i.e.
\begin{eqnarray}\label{ave2}
H^2 &=& {8\pi G\over3}\rho\,,\nonumber\\
\,\\
\dot H + H^2&=&-{4\pi G\over3}\left(3\mathcal{P}+\rho\right)\,,\nonumber
\end{eqnarray}
where both the total pressure and density are composed by pressureless matter, i.e. $\rho_m$, and by DE density, with
\begin{equation}
|\beta|=\left[\frac{1+\sqrt{(1-N_0-N_1)(1+N_0+N_1)}}{2}\right]^{1/2}.
\label{beta}
\end{equation}
The barotropic factor reads
\begin{equation}
\omega=\frac{-3N_0^2-2 N_0N_1(1+z)^2+N_1^2 (1+z)^4-12|\beta|^4}{3(N_0+N_1(1+z)^2)^2-12
 \Omega_m(1+z)^3 |\beta|^2+12 |\beta|^4}\,,
\label{omz}
\end{equation}
whose present value
\begin{equation}
  \omega_0=\frac{-2N_1(N_0+N_1)+3\left(1+\sqrt{1-\left(N_0+N_1\right)^2}\right)}{3\left(
   \Omega_m-1\right)\left(1+\sqrt{1-\left(N_0+N_1\right)^2}\right)}\,,
\end{equation}
which is compatible with present observations, when $N_0\ll1$ and $N_1\ll1$, showing that if entanglement occurred, the corresponding effects should be small. Despite small effects in the late time cosmology, entanglement is capable to drive the cosmic speed up with a negative pressure, alleviating the coincidence problem and providing a varying EoS.

\section{Conclusions}

In this paper, we relate the dynamical effects of DE to the hypothesis   of entanglement between cosmic quantum states. In doing so, we have shown that once quantum states are associated to  cosmological epochs, an entanglement process between them can be responsible of the observed cosmic speed up. The main features of this approach rely on two relevant facts. First, we assume that quantum effects can be detectable at present times. Second, we propose a mechanism to describe the DE density as the result of quantum interactions. The Entanglement States Ansatz (ESA)  enteres our picture, showing that it is possible to accelerate the universe without the need of a  cosmological constant acting as dark energy put by hand within the Einstein equations. We showed that this "coarse grained" approach to quantum cosmology is able to predict viable ranges of cosmological parameters and we demonstrated that cosmographic bounds can be  recovered in the framework of entangled cosmic states. We extended this picture by assuming that DE evolves in time, inferring barotropic corrections due to entanglement, once the source of DE entropy is the  Von Neumann entropy. In fact, once the entropy of the universe is associated to a varying Von Neumann entropy, the cosmological models evolve in time, showing that the $\Lambda$CDM model may represent nothing else but a limiting case of a more general theory. In addition, we proposed two possible cosmological models, by considering, first, an expanding adiabatic volume and second an apparent volume, scaling with $H^{-3}$. Finally, since quantum measurements characterize the weight of entanglement, we found a further cosmological model, choosing as measure of entanglement, the so called negativity parameter. Future developments could  characterize all the cosmological observables in terms of ESA, showing whether the role played by entanglement in present cosmology can reveal additional measurable effects. Nevertheless, the influence of ESA can  be related, in principle,  to different epochs of the universe evolution, i.e. inflation, nucleosynthesis, and so forth in order to understand if other detectable effects may occur tracing the cosmic evolution from quantum cosmology epochs up to now.

\end{document}